\documentclass[twocolumn,showpacs,preprintnumbers,amsmath,amssymb]{revtex4}

\usepackage{graphicx}
\usepackage{dcolumn}
\usepackage{bm}

\newcommand{\be}{\begin{equation}} \newcommand{\ee}{\end{equation}}
\newcommand{\bc}{\begin{center}} \newcommand{\ec}{\end{center}}
\newcommand{\bea}{\begin{eqnarray}} \newcommand{\eea}{\end{eqnarray}}

\begin{document}
\preprint{FT-031}

\title{A simple variational approach for an interacting\\
 Fermi trapped
 gas.\\}

\author{R. J\'auregui, R. Paredes and G. Toledo S\'anchez}

\affiliation{ Instituto de F\'{\i}sica, UNAM
A.~P.~20-364, M\'exico 01000 D.~F.~M\'exico\\}

\date{\today}

\begin{abstract}

Quantum states of a two-component Fermi trapped gas are described
by introducing an effective trap frequency, determined via
variational techniques. Closed expressions for the contribution of
a contact interaction potential to the total energy and the
pairing interaction are derived. They are valid for both few and
large number of particles, given the discrete nature of the
formulation, and therefore richer than the continuous expressions,
which are perfectly matched. Pairing energies within a shell are
explicitly evaluated and its allowed values at a given energy
level delimited. We show the importance of the interaction over
the trap energy as the number of particles ($N$) grows and the
temperature decreases. At zero temperature we find a polynomial
dependence of the interaction energy on the Fermi energy, whose
dominant term at large $N$ corresponds with the mean field
approximation result. In addition, the role of the strength of an
attractive potential on the total energy is exhibited.

\end{abstract}


\maketitle

\section{Introduction}
In the last years, trapped degenerate two-component Fermi atomic
gases have received great interest both from the theoretical
\cite{silvera,butts,houbiers,schneider,bruuna, bruunb} and the
experimental \cite{demarco,granade} point of view.  One of the
reasons is that these systems offer an ideal scenario to study
pairing interactions at variable interaction strength, density and
temperature. It has been recognized that atoms like $^{40}$K and
$^6$Li exhibit magnetically tunable Feshbach resonances so that
the $s$-wave scattering interaction may be varied from repulsive
to strongly attractive.  The accessible states for a dilute gas of
trapped atoms may be modeled in the zeroth-order approximation as
a system of noninteracting particles.  The simplest theoretical
approach for describing trapped Fermi atoms when the interactions
are not negligible, consists in approximating the system wave
function as properly symmetrized products of one-particle
eigenstates of an effective harmonic oscillator, whose frequency
is determined via a variational procedure. The purpose of this
work is to make an study of the predictions of such an approach.
In particular, we obtain closed expressions of the total
interaction energy that reflect its increasing relevance as the
number of particles in the system is augmented, as well as closed
expressions for pairing energies of atoms within a given shell.

\section{Formalism}
Let us consider a gas of Fermi atoms of mass $M$ with internal states
$\vert \uparrow\rangle$ and $\vert \downarrow\rangle$. They are in a
harmonic trap that for simplicity will be assumed to be of spherical
symmetry and characterized by a frequency $\omega_0$.  An asymmetric
harmonic trap can be treated in a completely analogous way.
We model the interaction between atoms in different internal
states by an effective contact potential
\begin{equation}
V(\vec x_{i^\downarrow} -\vec x_{j^\uparrow}) =
\frac{4\pi\alpha_0\hbar^2}{M}
\delta^{(3)}(\vec x_{i^\downarrow} -\vec x_{j^\uparrow})
\label{eq:s1}
\end{equation}
with $\alpha_0$ the scattering length. This potential is based on
scattering theory at low energies. It is inexact for at least two
reasons: it takes into account neither higher partial waves nor
the dependence of s-wave scattering on the momentum transfer. One
consequence of the latter is that no parameters beyond the
scattering length such as the effective range are included in
Eq.(\ref{eq:s1}). Thus, this potential is expected to be valid for
low energy transfer. In fact it has proven to be useful to
describe cold atoms at low densities \cite{houbiers} and it could
also be important for intermediate densities by introducing an
effective scattering length \cite{gehm}.

The Hamiltonian of the system is taken as
\begin{equation}
\hat H = \sum_{i}\big[\frac{\hat p_i^2}{2M}+\frac{1}{2}M\omega_0^2
\vert\vec x_i\vert^2\big] +\sum_{i^\downarrow,j^\uparrow} V(\vec
x_{i^\downarrow}-\vec x_{j^\uparrow}).
\label{eq:s2}
\end{equation}
We are interested in obtaining approximate eigenfunctions of this
Hamiltonian using the variational method. The validity of such an
approach is limited by the reliability of the model Hamiltonian.
As a trial wave function, we consider the product of two Slater
determinants, one for each internal state. These determinants are
built up by the one particle eigenfunctions of a spherical
harmonic oscillator of frequency $\omega^\prime$ and quantum
numbers $n_x,n_y$ and $n_z$. The value of $\omega^\prime$ should
reflect the interaction effects and in the limit of zero
interaction it becomes $\omega^\prime=\omega_0$. Notice, however,
that dramatic changes in $\omega^\prime$ with respect to
$\omega_0$ are incompatible with the assumptions supporting the
use of the contact interaction potential (\ref{eq:s1}).

Taking into account the results summarized in the
Appendix, the expectation value of the Hamiltonian for this trial
function is found to be
\begin{equation}
\langle \hat H \rangle = \hbar\omega_0\big[\frac{\epsilon_t
+\epsilon_0}{2} \big( a +\frac{1}{a}\big) +\epsilon_s a^{3/2}\big]
\end{equation}
where $a=\omega^\prime/\omega_0$,
\begin{eqnarray}
\epsilon_t
&=&\sum_{\{n_x^\uparrow,n_y^\uparrow,n_z^\uparrow\}}(n_x^\uparrow+n_y^\uparrow
+n_z^\uparrow)
+\sum_{\{n_x^\downarrow,n_y^\downarrow,n_z^\downarrow\}}(n_x^\downarrow+n_y^\downarrow
+n_z^\downarrow)\quad,\\ \epsilon_0 &=&
\frac{3}{2}(N^\uparrow+N^\downarrow),\\ \epsilon_s
&=&g\sum_{\{n_x^\uparrow,n_y^\uparrow,n_z^\uparrow\}}
\sum_{\{n_x^\downarrow,n_y^\downarrow,n_z^\downarrow\}}
I(n^\downarrow_x,n^\uparrow_x) I(n_y^\downarrow,n^\uparrow_y)
I(n_z^\downarrow,n^\uparrow_z).
\end{eqnarray}
with $N^{(i)}$ the number of particles in a given internal state $i$,
$g=\frac{4\pi\alpha_0}{\sqrt{\hbar / M\omega_0}}$ the scattering length in natural units
and
\begin{equation}\label{eq:inm}
I(n,m) = \frac{1}{\sqrt{2\pi}}\sum_{k=0}^{min\{n,m\}}
\frac{(2(m-k)-1)!!}{(2(m-k))!!}
\frac{(2(n-k)-1)!!}{(2(n-k))!!}\frac{(2k-1)!!}{(2k)!!}.
\end{equation}
$\epsilon_t$,  $\epsilon_0$ and $ \epsilon_s$ can be identified with the
trap, zero point and interaction energies in units of $\hbar \omega_0$,
respectively. Taking $a$ as a variational parameter, the condition that the
expectation value of the Hamiltonian becomes an extremum reduces to
\begin{equation}\label{eq:avar}
a^5
-\frac{1}{9}\Big(\frac{\epsilon_t+\epsilon_0}{\epsilon_s}\Big)^2(a^2-1)^2
=0.
\end{equation}
If the scattering length is negative, in order to admit a variational
solution corresponding to a minimum of the energy it is necessary that
\begin{equation}
\label{eq:acon}
a^{5/2} < -\frac{4(\epsilon_t+\epsilon_0)}{3\epsilon_s}.
\end{equation}

Once the proper value of the  variational parameter has been
determined, the properties of the system can be estimated.
In particular, the pairing energy between
two particles of quantum numbers $n^\uparrow$, $n^\downarrow$ whose
wave functions are $\phi_{n^\uparrow}(\vec x)$ and $ \phi_{n^\downarrow}(\vec x)$
is given by
\begin{eqnarray}
\langle V\rangle &=& \int d^3x \phi_{n^\uparrow}^*(\vec
x)\phi_{n^\uparrow}(\vec x) \phi_{n^\downarrow}^*(\vec
x)\phi_{n^\downarrow}(\vec x)\\ \nonumber
&=&\frac{4\pi\alpha_0}{\sqrt{\hbar /
M\omega^\prime}}I(n^\downarrow_x,n^\uparrow_x)
I(n_y^\downarrow,n^\uparrow_y) I(n_z^\downarrow,n^\uparrow_z).
\end{eqnarray}
For atoms in a trap with effective frequency $\omega^\prime$, an
 energy level $\epsilon_{{n_v}} = \hbar\omega^\prime (n_v +3/2)$
 exhibits a degeneracy $(n_v+2)(n_v+1)/2$ arising from the condition
 $n_x+n_y+n_z=n_v$.  Then the pairing energy between two particles in the
 same shell is directly given by the restriction
 $n_v^{\uparrow}=n_x^{\uparrow}+n_y^{\uparrow}+n_z^{\uparrow}
 =n_x^{\downarrow}+n_y^{\downarrow}+n_z^{\downarrow}=n_v^{\downarrow}$.
In order to illustrate the pairing energy features, let us explore
the behavior of the universal dimensionless quantity
\begin{equation}
\delta_{n_v}(n_x,n_y,n_z) = I(n_x,n_x)I(n_y,n_y)I(n_z,n_z), \quad\quad
n_x+n_y+n_z =n_v.
\end{equation}
 For a given quantum number $n$, the integrals $I(n,m)$ achieve its
 maximum value for $n=m$.
In Figure 1, we plot the allowed region of the $\delta_{n_v}$ values,
delimited by solid line curves. They correspond to its maximum $\delta_{max}$ (top) and
minimum $\delta_{min}$ (bottom) values, for $0<n_v<200$.
Systematically $\delta_{max}$ corresponds to two quantum numbers equal
to zero, for instance, $n_y=0$, $n_z=0$, while $n_x=n_v$.  Meanwhile
$\delta_{min}$ is achieved when $n_v$ is decomposed, $e.$ $g.$, in the
form $n_x\sim (3/5) n_v$ and $n_y=n_z\sim 1/5 n_v$.

\section{Ground State.}
As a particular application of these formulae we consider the
ground state wave function corresponding to a configuration where
all the possible states up to the non interacting Fermi energy
$\epsilon_F = \hbar\omega_0({\cal M}_F +3/2)$ are occupied. To
characterize the system under these conditions, we need to
calculate the summations
\begin{eqnarray}
\sum_{n_x=0}^{{\cal M}_F}\sum_{n_y=0}^{{\cal
M}_F-n_x}\sum_{n_z=0}^{{\cal M}_F-n_x-n_y} 1 &=&\frac{({\cal M}_F+1)^3
+3 ({\cal M}_F+1)^2 +2({\cal M}_F+1)}{6}\\ \sum_{n_x=0}^{{\cal
M}_F}\sum_{n_y=0}^{{\cal M}_F-n_x}\sum_{n_z=0}^{{\cal M}_F-n_x-n_y}
n_x &=&\frac{({\cal M}_F+1)^4 +2({\cal M}_F+1)^3 -({\cal M}_F+1)^2
-2({\cal M}_F+1)}{24}.
\end{eqnarray}
The first sum let us evaluate the number of particles $N^{(i)}$ for
 such a configuration in terms of the Fermi energy. While the second
 gives one third of the corresponding trap energy $\epsilon_t^{(i)}$,
 in $\hbar\omega_0$ units, neglecting the zero point energy.
 Notice that for  large ${\cal M_F}$ the trap energy becomes
\begin{equation}
\epsilon_t^{(i)} = \frac{3}{4} N^{(i)}{\cal M_F} = \frac{3}{4}
N^{(i)}\big[\frac{\epsilon_F}{\hbar\omega_0} -\frac{3}{2}\big].
\end{equation}
That is, the discrete summation relation between these variables
coincides with that obtained in the standard procedure that works
with an approximate continuum spectra.  In Table I, the behavior
of the dimensionless quantities $N^{(i)}=N^\uparrow=N^\downarrow$,
$\epsilon_t^{(i)}$ and $\tilde\epsilon_s= \epsilon_s/g$ as a
function of ${\cal M}_F$ is illustrated.  Notice the increasing
relevance of the interaction term $\tilde\epsilon_s$ compared to
the $\epsilon_t$. While at ${\cal M_F}=10$ it represents only a
$13.6\%$, at ${\cal M_F}=300$ it becomes $47\%$ as big as $\epsilon_t$.
In fact, for $0<{\cal M}_F\sim 200$, we have found that
$\tilde\epsilon_s$  has a polynomial behavior of the form
\begin{equation}
\tilde\epsilon_s = (2\pi)^{-3/2}+ {\cal M}_F^{1/2}(c_0+c_1{\cal
M}_F+c_2{\cal M}_F^2 +c_3{\cal M}_F^3+c_4{\cal M}_F^4)
\label{eq:int}
\end{equation}
with
\begin{eqnarray}
c_0&=&0.11671733,\quad\quad c_1= 0.18747855,\quad\quad
c_2=0.12399665,\nonumber\\ c_3&=&0.03706759,\quad\quad c_4= 0.00411862 \nonumber
\end{eqnarray}
with a relative error of the order of $10^{-7}$ for ${\cal M}_F>40$
and of order less or equal to $10^{-4}$ for ${\cal M}_F<40$.  That is,
our numerical calculations show that the interaction energy scales as
${\cal M}_F^{4.5}$ for large ${\cal M}_F$ growing faster than
$\epsilon_t\sim {\cal M}_F^4$. This numerical result can be
understood within the mean field approximation as follows:

In general, for a given mixture of Fermi atoms described by the
product of two Slater determinants formed by one-particle
eigenfunctions $\phi^{(i)}_a(\vec r_j)$ of an effective one particle
Hamiltonian, $\hat h = \hat p^2/2M + v(\vec r)$,
the expectation value of a contact interaction potential is determined
by
\begin{eqnarray}\label{eq:ctct}
\sum_{i^\downarrow,j^\uparrow} \langle \delta(\vec
x_{i^\downarrow}-\vec x_{j^\uparrow})\rangle &=&
\sum_{n^\uparrow,n^\downarrow}\int d^3x \phi_{n^\uparrow}^*(\vec
x)\phi_{n^\uparrow}(\vec x) \phi_{n^\downarrow}^*(\vec
x)\phi_{n^\downarrow}(\vec x)\\ &=&\int d^3x\rho^\uparrow(\vec
x)\rho^\downarrow(\vec x)
\end{eqnarray}
where $\rho^{(i)}(\vec x)$ represents the spatial density of atoms of
{\it i}-type.  In our problem, the one particle Hamiltonian is that of a
spherical harmonic oscillator of frequency $\omega^\prime$. Then, at
zero temperature and for a large number of atoms, the Thomas-Fermi
approximation leads to the expression \cite{silvera,butts}
\begin{equation}
\rho^{(i)}(\vec x)_{T=0}\sim
\frac{N^{(i)}}{R^3_F}\frac{8}{\pi^2}\Big[1-\frac{r^2}{R^2_F}\Big]^{3/2}
,\quad\quad R_F=\Big[\frac{2\epsilon_F}{M\omega^{\prime
2}}\Big]^{1/2}.
\end{equation}
So that,
\begin{eqnarray}
\frac{4\pi\alpha_0\hbar^2}{M} \sum_{i^\downarrow,j^\uparrow}
\langle\vert \delta(\vec x_{i^\downarrow}-\vec
x_{j^\uparrow})\vert\rangle &\sim& g \hbar \omega^\prime
\frac{512}{\sqrt{2}\pi^32835}
\Big[\frac{\epsilon_F}{\hbar\omega^\prime}\Big]^{4.5}\\ &\sim&
0.004118625 \hspace{.2cm} g \hbar\omega_0 a^{3/2}{\cal M}_F^{4.5}
\end{eqnarray}
Eq.~(\ref{eq:int}) is consistent with the later result and, in fact,
generalizes this formula for small ${\cal M}_F$.

Another useful result to exploit the variational approach, for large
$N^{(\uparrow)}=N^{(\downarrow)}=N^{(i)}$, is that
the variational equation (\ref{eq:avar}) can be approximated  by
\begin{equation}
0.419372 g {\tilde R}_F a^5 - (a^2-1)^2 = 0 ,
\end{equation}
where ${\tilde R}_F=R_F/\sqrt{\hbar/M\omega^\prime}= (48
N^{(i)})^{1/6}$ is the characteristic size of the trapped
degenerate Fermi gas in the natural units of the trap.
In Table II, the results of the variational calculation are
exemplified for ${\cal M}_F=150$, corresponding to about half a
million atoms of a given internal state.  An attractive
interaction is considered and the corresponding coupling constant
$g$ is varied, leading to optimal values of $a$ via the
variational equation. Notice that the information about the trap
parameter ($\omega_0$) and the atom characteristics ($M$ and
$\alpha_0$) are contained in $g$. Therefore the results are not
restricted to neither a specific atom nor trap. In this table, the
exact expectation values of the Hamiltonian in $\hbar\omega_0$
units divided by the total number of particles
$N=N^\uparrow+N^\downarrow$ are shown for $\omega^\prime
=\omega_0$ ( $\langle \tilde E\rangle_{\omega^\prime =\omega_0}$)
and for an optimized choosing of the variational parameter $a$ (
$\langle \tilde E\rangle_{min}$).  As predicted by
Eq.~(\ref{eq:acon}) no minima can be found for $\vert g\vert
>0.9$. In this case the expectation value of the energy
exhibits a maximum at a given value of $a<1$ and decreases
monotonically for $a>1$. That is, the gas would behave as trapped
in a harmonic potential of arbitrarily large frequency. Therefore,
it would collapse. However, this prediction is beyond the validity
region of the model.

A mechanical instability of a strongly interacting Fermi gas
arises when attractive interactions overcome repulsive effects
such as the Fermi pressure \cite{houbiers}. In actual experimental
realizations \cite{gehm}, there are mechanisms not represented by
the simple model Eq.(\ref{eq:s1}) that may stabilize the system.
For instance, repulsive contributions to the
interaction arising from higher order partial waves \cite{roth}, or within
the s-wave scattering scheme, effective potentials
resulting from adding the ladder diagrams in a many-body
perturbative approach \cite{heiselberg}. Assuming densities that
guarantee an interatomic separation larger than the range of the
potential but much shorter than the scattering length, Heiselberg
found that an effective s-scattering amplitude arises, which always
favors the stability of a two component Fermi gas.

A mixture of spin-polarized
atomic $^6$Li in two possible hyperfine states constitutes a
specially interesting system \cite{houbiers,granade,gehm}. It
 exhibits magnetic tunable
Feshbach resonances that enable the variation of the scattering
length from strongly repulsive to strongly attractive.  In fact,
an anomalously large and negative scattering length $\alpha_0 =
-2160$ a$_0$ has been already measured \cite{abraham}. The
coupling constant corresponding to this value of $\alpha_0$, for a
typical trap of mean frequency $\omega_0\sim 3$kHz is $g=-0.725$.
Then the effective frequency for half a million atoms is
$\omega^\prime = 1.48\omega_0$. It differs significantly from the
trap frequency; the model potential and our variational approach
are at the border of its applicability.

\begin{table}
\caption{\label{tab:table1}Behavior of the dimensionless
quantities: number of particles N, trap energy $\epsilon_t$ and
interaction energy $\tilde\epsilon_s=\epsilon_s/g$ as a function
of ${\cal M}_F$. }
\begin{ruledtabular}

\begin{tabular}{||r|r|r|r||}
 ${\cal M}_F$ & $N^{(i)}$ & $\epsilon_t^{(i)}$ &
$\tilde\epsilon_s$\cr\hline\hline 5 & 56& 210 & 25.4 \cr\hline 10 &
286& 2145 & 293.0 \cr\hline 20 & 1771& 26565 & 4512.3 \cr\hline 30 &
5456& 122760 & 24396.9 \cr\hline 40 & 12341& 370230 & 82991.0
\cr\hline 50 & 23426& 878475 & 217041.8 \cr\hline 60 & 39711& 1786995
& 479024.0 \cr\hline 70 & 62196& 3265290 & 938926.7 \cr\hline 80 &
91881& 5512860 & 1685871.8 \cr\hline 90 & 129766& 8759205 & 2829605.9
\cr\hline 100 & 176851& 13263825 & 4501889.6 \cr\hline 150 & 585276&
65843550 & 27103302.6 \cr\hline 300 & 1373701& 206055150 & 97458248.4
\cr
\end{tabular}
\end{ruledtabular}
\end{table}

\begin{table}
\caption{\label{tab:table2} Average energy per atom for
$\omega^\prime=\omega_0$, $\langle \tilde E\rangle_{\omega^\prime
=\omega_0}/N^{(i)}$ and for an optimized variational parameter
$a$, $\langle \tilde E\rangle_{min}/N^{(i)}$, as a function of the
strength of an attractive potential, $g$ . }
\begin{ruledtabular}

\begin{tabular}{||c|c|c|c||}
$g$ &$\langle \tilde E\rangle_{\omega^\prime
=\omega_0}/N^{(i)}$& $a$& $\langle \tilde
E\rangle_{min}/N^{(i)}$\cr\hline\hline -0.05 & 111.34 & 1.02
&111.33\cr
\hline -0.10 & 110.18 & 1.03 &110.13\cr
\hline -0.20& 107.87 & 1.07 &107.63\cr
\hline -0.30 & 105.55 & 1.11 &104.99\cr
\hline-0.40 & 103.24 & 1.17 &103.24\cr
\hline -0.50 & 100.92 & 1.23 & 99.12\cr
\hline-0.60 & 98.61 & 1.32 & 95.79\cr
\hline-0.7251& 95.71 & 1.48 & 91.03\cr
\hline -0.80 & 93.98 & 1.65 & 87.64\cr
\hline -0.85 & 92.82 & 1.90 & 84.93\cr

\end{tabular}
\end{ruledtabular}
\end{table}

\section{Interaction energy for $T\ne 0$.}
A detailed description of the behavior of the Fermi gas at finite
temperature including interactions, can be given by evaluating the
variational wave functions for a set of accessible quantum states and
performing the proper statistical average to infer properties like
average energies or densities.  In this section we shall make an
approximate calculation where the starting point is
Eq.~(\ref{eq:ctct}).  This approach $has$ $not$ a rigorous theoretical
basis but should give us a reasonable estimate of the energy shift at
finite temperatures.  According to the Thomas-Fermi approximation,
valid for a large number of particles, the atomic density for a
trapped Fermi gas, at a given temperature $T$, is \cite{butts}
\begin{equation}
\rho^{(i)}(\vec x,T) =\frac{1}{2\pi}
\int\frac{d^3k}{e^{\beta[\hbar\vert\vec k\vert^2/2M + 1/2
M\omega^{\prime 2}r^2]-2}+1},
\end{equation}
where $\beta=1/k_B T$. Defining the function
\begin{equation}
{\cal I}(\xi)=\int d\eta\int d\eta^\prime\int d\zeta\eta^2\eta^{\prime
2}\zeta^2\big[e^{\eta^2+\zeta^2-\xi}+1\big]^{-1} \big[e^{\eta^{\prime
2}+\zeta^2-\xi}+1\big]^{-1},
\end{equation}
we can  estimate the interaction energy at such temperature
by generalizing equation (\ref{eq:ctct}) to consider temperature
dependent densities:
\begin{eqnarray}
\langle V(N,T)\rangle&=& \frac{4\pi\alpha_0\hbar^2}{M}\int d^3x
\rho_\uparrow(\vec x,T)\rho_\downarrow(\vec x,T)\\
&=&\frac{4\pi\alpha_0\sqrt{M}2^{9/2}}{\pi^3
\hbar^4\omega^{\prime 3}\beta^{9/2}}{\cal I}(\xi)
\end{eqnarray}
where the parameter $\xi=\mu\beta$ is given implicitly by
\begin{equation}
N^{(i)}=\frac{1}{2\hbar\omega^\prime}
\int\frac{\epsilon^2d\epsilon}{e^{\beta(\epsilon-\mu)}+1}
\end{equation}
with $\mu$ the chemical potential.

In Figure 2, the behavior of the dimensionless quantity $\tilde v(N,T)
=\langle V(N,T)\rangle \sqrt{\hbar\omega^\prime M}/(4\pi \alpha_0 N)$
as a function of the temperature is exemplified for $N=10^5$. We can
observe that  although the interaction may be negligible for
$k_B T>\epsilon_F$ its relevance increases dramatically for
$k_B T<\epsilon_F$.

\section{Discussion and conclusions.}

We have studied the role of the $s-$wave contact interaction between
Fermi atoms in different states confined in a trap by using a simple
variational approach, where the interaction effects on the variational
wave function are resumed in an effective frequency.
We have found closed expressions for the overlap integrals that admit
a fast convergence and allow a quite direct
evaluation of pairing energies and total interaction energies.

Concerning the pairing energies between atoms with the same trap
energy $\epsilon_T =\hbar\omega^\prime( n_v +3/2)$ we
determine the region of relevance: they become
maximum for atoms with equal quantum numbers and in states which
oscillate in a given axis with excitation number $n_i=n_v$ while
remain in the minimum oscillating state in the other directions $n_j =
0$ $j\ne i$. For a spherical trap, there is no privileged axis of
oscillation.  Meanwhile, a direct extension of the results here
obtained, shows that for a completely asymmetric trap, oscillations in
the direction of the highest effective frequency exhibit a higher
pairing energy.  These results can be compared with those found
for the isotropic trap in spherical coordinates \cite{heiselberg2}.
There the individual wave functions have a well defined
angular momentum $l$, and it is natural to consider paired
state functions within a given
shell $n_v$ that are eigenfunctions of the total angular momentum $L$.
In that case the greatest pairing energies
correspond to $n_v=l$ and $L=0$.  These paring energies $do$ $not$
decrease as $n_v$ increases.  For our much simpler two-particle
states, $\delta_{max}$ decreases as $n_v$. However, our results
have a direct translation for anisotropic traps, while the
relevance of angular momentum coupled states is highly dependent on
the spherical symmetry.  Thus, it is important to search for symmetry
operators that could permit the definition of more tightly bound
two-particle states for anisotropic traps.

We have found that  the contribution of the contact interaction
potential to the total energy of the fermionic ground state
behaves as a polynomial, Eq.~\ref{eq:int}. Being valid for both
few and many atoms, this expression shows the increasing relevance
of the interaction terms with respect to the trap energies as the
number of particles grows. In the large $N$ region, the
variational equation to determine the effective trap frequency
relates the variational parameter with the geometric factor
$\tilde R_F$. Meanwhile the stability criteria (\ref{eq:acon})
restricts the admissible values of the coupling constant $g$.
Beyond this limit both the model potential and the simple trial
wave function have not a clear justification. As an extension to
finite temperature we elaborate on a possible form of the
interaction energy expectation value and find that it becomes
increasingly relevant  for $k_B T<\epsilon_F$ as $T$ diminishes.

The approximations made in this work could be valid not just for
dilute gases but also for denser systems by taking $\alpha_0$ as
an effective scattering length. A more precise study can be made
by solving the Hartree-Fock equations of the system using the
contact interaction \cite{bruuna} or working with a different
model potential and more sophisticated trial wave functions
\cite{carlson}. The computational load of such an approach
restricts actual calculations to no more than hundreds of
particles.  Here, we have shown explicit results up to $10^6$
particles.  The possibility of incorporating correlation effects,
that is, working with more than one Slater determinant for each
type of atoms as well as other forms of the effective interaction
potential is under way.

\begin{acknowledgments}
This work was partially supported by Conacyt M\'exico under grant 41048-A1,
and we thank Prof. H.~Vucetich for useful discussions.
\end{acknowledgments}

\section{Appendix.}
In this appendix, integrals of the form
\begin{equation}
{\cal H}(n,m)= \int_{-\infty}^\infty d\zeta
e^{-2\zeta^2}H_n^2(\zeta)H_m^2(\zeta).
\end{equation}
with $H_n$ the Hermite polynomials are evaluated.  A closed expression of
these integrals may be obtained employing the Rodrigues formula for
Hermite polynomials.  This leads to expressions that contain
summations of terms with alternating signs.  Here, we look for an
expression in terms of finite fast convergent summations.  To that
end, consider the fact \cite{grashteyn}
\begin{equation}
[H_n(\zeta)]^2 = \frac{(-1)^n}{\pi}\frac{(n!)^22^n}{(2n)!}\int_0^\pi
(cos y)^nH_{2n}(x(1-sec y)^{1/2}).
\end{equation}
Besides \cite{prudnikov}
\begin{eqnarray}
\int_{-\infty}^{\infty}e^{-px^2}H_{2m}(bx)H_{2n}(cx)dx&=&
(2m)!(2n)!\sqrt{\frac{\pi}{p}} \sum_{k=0}^{min\{n,m\}}
\frac{1}{(m-k)!(n-k)!(2k)!}\cdot\\ \nonumber &\cdot&\big(\frac{b^2
-p}{p}\big)^{m-k} \big(\frac{c^2
-p}{p}\big)^{n-k}\big(\frac{2bc}{p}\big)^{2k}
\end{eqnarray}
so that
\begin{eqnarray}
{\cal H}(n,m)&=&\frac{(n!)^2(m!)^2}{\pi^2}
\sqrt{\frac{\pi}{2}}\sum_{k=0}^{min\{m,n\}}
\frac{(-2)^{2k}}{(m-k)!(n-k)!(2k)!}  \cdot \\ \nonumber
&\cdot&\int_0^\pi dy (\cos y + 1)^{n-k}(\cos y-1)^k \int_0^\pi
dy^\prime (\cos y + 1)^{m-k}(\cos y-1)^k
\end{eqnarray}
but
\begin{equation}
\int_0^\pi dy (\cos y+1)^{m-k}(\cos y -1)^k = (-1)^k 2^m\int_0^\pi
\cos^{2(m-k)}(y/2)\sin^{2k}(y/2)
\end{equation}
and \cite{grashteyn}
\begin{equation}
\int_0^{\pi/2}\sin^{\mu-1} x \cos^{\nu-1} x dx = \frac{1}{2}
B(\frac{\mu}{2},\frac{\nu}{2})
\end{equation}
with $B(r,s)$ the Bernoulli numbers, thus
\begin{eqnarray}
{\cal H}(n,m)&=&\frac{2^{m+n}(n!)^2(m!)^2} {\pi^{3/2}\sqrt{2}}\cdot \\
\nonumber
&\cdot&\sum_{k=0}^{min\{m,n\}}\frac{4^kB(m-k+1/2,k+1/2)B(n-k+1/2,k+1/2)}{(m-k)!(n-k)!(2k)!}.
\end{eqnarray}
Besides \cite{grashteyn}
\begin{equation}
B(r,s) = \frac{\Gamma(r)\Gamma(s)}{\Gamma(r+s)},\quad\quad
\Gamma(n+1/2) =\frac{\sqrt{\pi}}{2^n}(2n-1)!!
\end{equation}
We can thus conclude that
\begin{equation}
{\cal H}(n,m) = n!m!\sqrt{\frac{\pi}{2}}\sum_{k=0}^{min\{m,n\}}
\frac{4^k(2m-2k-1)!!(2k-1)!!(2n-2k-1)!!(2k-1)!!}{(m-k)!(n-k)!(2k)!}
\end{equation}
and the expectation value of the one-dimensional contact interaction
is then given by
\begin{eqnarray}
\langle \delta(x^\uparrow - x^\downarrow)\rangle
&=&\sqrt{\frac{M\omega}{\hbar}}\frac{1}{\sqrt{\pi} 2^n n!}\frac{1}{\sqrt{\pi} 2^m m!}{\cal H}(n,m) \nonumber\\
&=&\frac{1}{\sqrt{2\pi}}\sqrt{\frac{M\omega}{\hbar}}
\sum_{k=0}^{min\{m,n\}}
\frac{(2n-2k-1)!!}{(2n-2k)!!}\frac{(2m-2k-1)!!}{(2m-2k)!!}\frac{(2k-1)!!}{(2k)!!}\nonumber\\
&=&\sqrt{\frac{M\omega}{\hbar}}I(n,m).
\end{eqnarray}

\newpage

\begin{figure}
\includegraphics{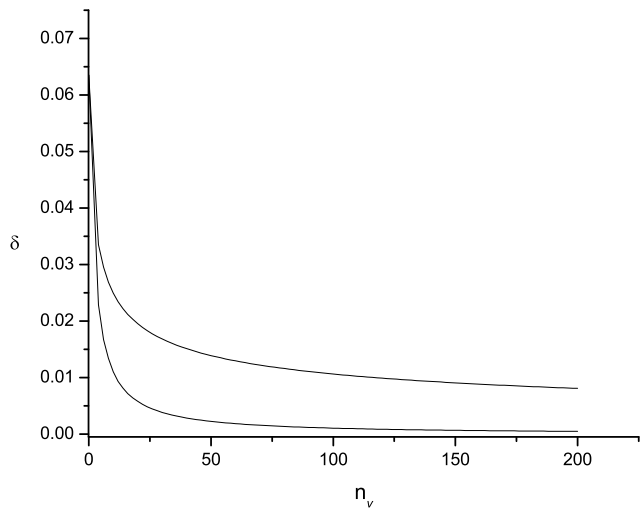}
\caption{Behavior of the dimensionless pairing interaction $\delta_{n_v}$ as a
function of the energy level quantum number ${n_v}$. The allowed region is delimited
by its maximum $\delta_{max}$ (top curve) and minimum  $\delta_{min}$ (bottom curve).}
\end{figure}

\begin{figure}
\includegraphics{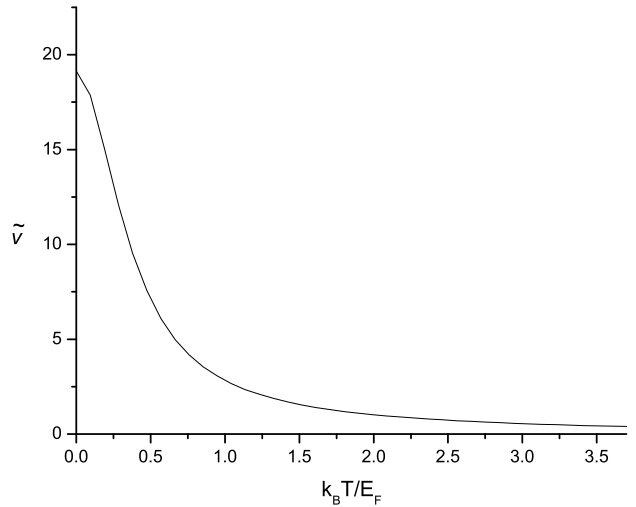}
\caption{Behavior of the interaction $\tilde v(N,T)$ for $N=10^5$ atoms as a function of the temperature.}
\end{figure}

\end{document}